\documentclass[aps,prl,twocolumn,showpacs,superscriptaddress]{revtex4}

\usepackage{graphicx}

\begin{document}


\title{First Use of High Charge States for Mass Measurements of Short-lived Nuclides in a Penning Trap}

\author{S.~Ettenauer}
  \email[Corresponding author: ]{sette@triumf.ca}
  \affiliation{TRIUMF, 4004 Wesbrook Mall, Vancouver, BC V6T 2A3, Canada}
  \affiliation{Department of Physics and Astronomy, University of British Columbia, Vancouver, BC V6T 1Z1, Canada}
  
 \author{M.~C.~Simon}
  \affiliation{TRIUMF, 4004 Wesbrook Mall, Vancouver, BC V6T 2A3, Canada}
  
\author{A.~T.~Gallant}
  \affiliation{TRIUMF, 4004 Wesbrook Mall, Vancouver, BC V6T 2A3, Canada}
    \affiliation{Department of Physics and Astronomy, University of British Columbia, Vancouver, BC V6T 1Z1, Canada}
  
\author{T.~Brunner}
  \affiliation{TRIUMF, 4004 Wesbrook Mall, Vancouver, BC V6T 2A3, Canada}
  \affiliation{Physik Department E12, Technische Universit\"at M\"unchen, D-85748 Garching, Germany}

\author{U.~Chowdhury}
   \affiliation{TRIUMF, 4004 Wesbrook Mall, Vancouver, BC V6T 2A3, Canada}
   \affiliation{Department of Physics and Astronomy, University of Manitoba, Winnipeg, MB R3T 2N2, Canada}

\author{V.~V.~Simon}
  \affiliation{TRIUMF, 4004 Wesbrook Mall, Vancouver, BC V6T 2A3, Canada}
  \affiliation{Max-Planck-Institut f\"ur Kernphysik, Saupfercheckweg 1, 69117 Heidelberg, Germany}
  \affiliation{Ruprecht-Karls-Universit\"at, Heidelberg, Germany}
  
\author{M.~Brodeur}
  \affiliation{TRIUMF, 4004 Wesbrook Mall, Vancouver, BC V6T 2A3, Canada}
  \affiliation{Department of Physics and Astronomy, University of British Columbia, Vancouver, BC V6T 1Z1, Canada}
  \affiliation{National Superconducting Cyclotron Laboratory, Michigan State University, East Lansing, MI 48824, USA}

\author{A.~Chaudhuri}
    \affiliation{TRIUMF, 4004 Wesbrook Mall, Vancouver, BC V6T 2A3, Canada}
   
\author{E.~Man\'{e}}
  \affiliation{TRIUMF, 4004 Wesbrook Mall, Vancouver, BC V6T 2A3, Canada}
  
\author{C.~Andreoiu}
  \affiliation{Department of Chemistry, Simon Fraser University, Burnaby, BC V5A 1S6, Canada}

\author{G.~Audi}
  \affiliation{CSNSM-IN2P3-CNRS, Universit\'e Paris 11, 91405 Orsay, France}

\author{J.~R.~Crespo L\'{o}pez-Urrutia}
  \affiliation{Max-Planck-Institut f\"ur Kernphysik, Saupfercheckweg 1, 69117 Heidelberg, Germany}

\author{P.~Delheij}
   \affiliation{TRIUMF, 4004 Wesbrook Mall, Vancouver, BC V6T 2A3, Canada}

\author{G.~Gwinner}
  \affiliation{Department of Physics and Astronomy, University of Manitoba, Winnipeg, MB R3T 2N2, Canada}
  
\author{A.~Lapierre}
    \affiliation{TRIUMF, 4004 Wesbrook Mall, Vancouver, BC V6T 2A3, Canada}
   \affiliation{National Superconducting Cyclotron Laboratory, Michigan State University, East Lansing, MI 48824, USA}

\author{D.~Lunney}
   \affiliation{TRIUMF, 4004 Wesbrook Mall, Vancouver, BC V6T 2A3, Canada}
  \affiliation{CSNSM-IN2P3-CNRS, Universit\'e Paris 11, 91405 Orsay, France}
  
\author{M.~R.~Pearson}
  \affiliation{TRIUMF, 4004 Wesbrook Mall, Vancouver, BC V6T 2A3, Canada}
  
\author{R.~Ringle}
    \affiliation{National Superconducting Cyclotron Laboratory, Michigan State University, East Lansing, MI 48824, USA}
  
\author{J.~Ullrich}
  \affiliation{Max-Planck-Institut f\"ur Kernphysik, Saupfercheckweg 1, 69117 Heidelberg, Germany}

\author{J.~Dilling}
  \affiliation{TRIUMF, 4004 Wesbrook Mall, Vancouver, BC V6T 2A3, Canada}
   \affiliation{Department of Physics and Astronomy, University of British Columbia, Vancouver, BC V6T 1Z1, Canada}
  
\date{\today}

\begin{abstract}
Penning trap mass measurements of short-lived nuclides have been performed for the first time with highly-charged ions (HCI), using the TITAN facility at TRIUMF. Compared to singly-charged ions, this provides an improvement in experimental precision that scales with the charge state $q$. Neutron-deficient Rb-isotopes have been charge bred in an electron beam ion trap to $q = 8 - 12+$ prior to injection into the Penning trap. In combination with the Ramsey excitation scheme, this unique setup creating low energy, highly-charged ions at a radioactive beam facility opens the door to unrivalled precision with gains of 1-2 orders of magnitude. The method is particularly suited for short-lived nuclides such as the superallowed $\beta$ emitter $^{74}$Rb ($T_{1/2}=65$ ms). The determination of its atomic mass and an improved $Q_{EC}$-value are presented.
\end{abstract}


\maketitle
\begin{figure}
\includegraphics[height=.21\textheight]{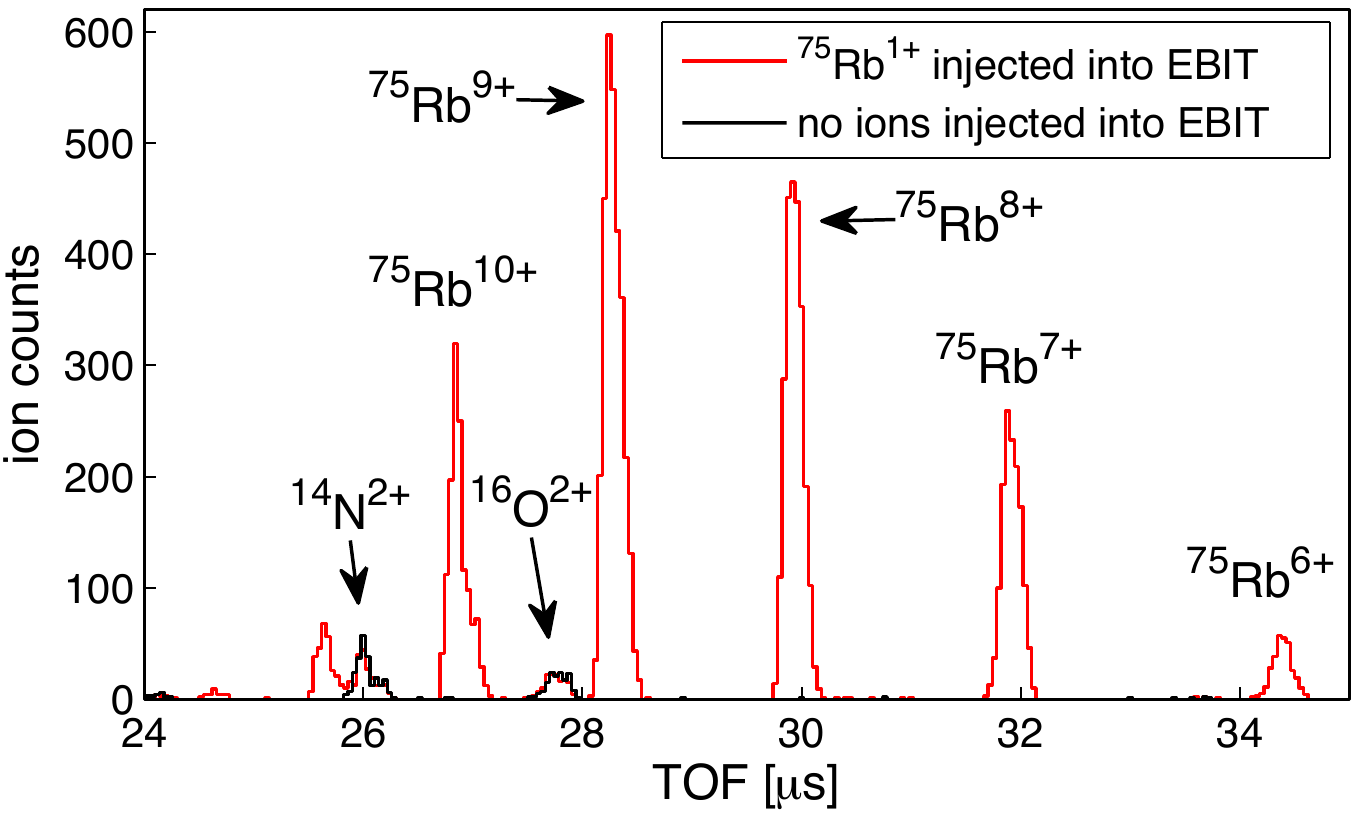}
\caption{\label{fig:chargeDistributionTOF}(color online) TOF spectrum of 500 ion bunches of radioactive $^{75}$Rb extracted from the EBIT with an 800 ns extraction pulse after 35 ms of charge breeding with a 10 mA, 2.5 keV electron beam.  $^{16}$O$^{2+}$ and $^{14}$N$^{2+}$ are due to ionized and further charge-bred residual gas in the EBIT. }
\end{figure}

Since their introduction into rare isotope research over twenty years ago \cite{springerlink:10.1007/BF02394875,PhysRevLett.65.3104}, Penning traps have made major contributions to the exploration of the nuclear mass surface as evidenced by the large number of existing and proposed facilities \cite{Blaum20061} as well as the wealth of experimental results \cite{springerlink:10.1007/s10751-009-0031-8}. Advances in experimental techniques now allow measurements for virtually all low energy, rare isotope beams as Penning traps have been able to access nuclides with half-lives below 10 ms \cite{PhysRevLett.101.202501} as well as superheavies with production yields of less than 1 particle per second \cite{BlockNature463.785}. The widespread success of Penning traps is due to their precision following 
\begin{equation}
\frac{\delta m}{m}\propto\frac{m}{qBT_{\rm rf}\sqrt{N_{\rm ion}}}
\label{eq:PenningTrapPrecision}
\end{equation}
\cite{Bollen20013}, where $\delta m/m$ is the achievable relative precision in mass $m$, $q$ is the ion's charge state and $B$ is the magnetic field strength. The measurement time $T_{\rm rf}$ and the number of ions $N_{\rm ion}$ are limited by a nuclide's half-life and possibly by its production yield at radioactive beam facilities and efficiency of the spectrometer. Measurements are generally performed with singly-charged ions (SCI) or in special cases, where coupled to a gas stopper cell, with $q=2+$. Penning trap mass studies utilizing highly-charged ions (HCI) have been successfully pioneered with stable nuclides \cite{PhysRevLett.83.4506}. Here  the requirements of high efficiency and short measurement times are less relevant compared to working with radioactive ions. In the realm of rare isotope science with Penning traps, HCI represent a thus far unexplored opportunity to improve the experimental precision further circumventing constraints imposed by short half-lives and lower yields when probing the limits of nuclear existence. 
\newline
The superallowed $\beta$ emitter $^{74}$Rb is a prime example where a short half-life of only 65~ms poses a real challenge to experiment. Despite several Penning trap mass measurements \cite{springerlink:10.1140/epja/i2001-10216-x,PhysRevLett.93.072502,PhysRevC.76.045504}, the total transition energy, $Q_{EC}$,  still contributes significantly to the uncertainty of its corrected $\mathcal{F}t$-value, only surpassed by theoretical uncertainties of the isospin-symmetry breaking corrections $\delta_C$ \cite{PhysRevC.79.055502}. These have  recently been reduced by experimentally providing the $^{74}$Rb rms charge radius as an input for the calculation of $\delta_C$  \cite{ManeChargeRadiusRb74}. The $Q_{EC}$-value and $\delta_c$ are now close to sharing the same weight to the total uncertainty of the $\mathcal{F}t$-value.  Among all superallowed $\beta$ emitters used to extract $V_{\rm ud}$ of the Cabibbo-Kobayashi-Maskawa (CKM) matrix \cite{PhysRevC.79.055502}  $^{74}$Rb has the highest atomic number, $Z$. It is hence of special importance in attempts to distinguish between conflicting nuclear models since $\delta_C$ approximately scales as $Z^2$ \cite{Grinyer2010236,PhysRevC.82.065501}. In this letter we present the first Penning trap mass measurements of short-lived HCI performed, with TRIUMF's Ion Trap for Atomic and Nuclear science (TITAN) \cite{Dilling2006198}, including a successful mass determination of  $^{74}$Rb$^{8+}$. 
\newline
Neutron-deficient Rb isotopes were produced at TRIUMF's ISAC facility \cite{dombsky:978} by bombarding a Nb-target with a 98 $\mu$A, 500 MeV proton beam from the cyclotron. The surface-ionized Rb beam was accelerated to 20 keV and mass separated prior to injection into TITAN's radio-frequency quadrupole (RFQ) cooler and buncher \cite{TITAN-RFQ} where the ions were accumulated and cooled through collision with a He buffer gas. Extracted ion bunches were transferred with a beam energy of about 2~keV into the electron beam ion trap (EBIT)\cite{Lapierre201054}. Operated with a 10 mA, 2.5 keV electron beam, the EBIT confined the ions radially by the space-charge of the electron beam and an axial magnetic field of 3 T. The central, trapping drift tube was biased (at $U_{\rm trap}$) slightly below the beam energy to remove most of the ions' kinetic energy. To provide confinement in the axial direction, a bias voltage of $\approx$100 V above the central drift tube was applied to the neighbouring drift tubes, one of which was lowered during the ions' capture and extraction. Through electron-impact ionization the initially singly charged ions were charge bred to higher charge states. Due to a kinetic energy of $q\cdot U_{\rm trap}$ after extraction from the EBIT, different charge states $q_i$ can be identified via time of flight (TOF) as illustrated in Fig.\ref{fig:chargeDistributionTOF}.
At a fixed electron beam setting, a certain charge state can be maximized in abundance by optimizing the charge breeding time. For instance, the number of Rb ions with $q=8+$ reached its maximum at a flat plateau of $\approx$20-27 ms of breeding. A Bradbury-Nielson ion gate (BNG)\cite{EXO-TITAN-BNgate} allowed the selection of one charge state by opening the gate for 300-500~ns during the beam transport from the EBIT to TITAN's measurement Penning trap (MPET). In the MPET, the ion's cyclotron frequency $\nu_c=qB/(2\pi m)$ was determined by the time-of-flight ion-cyclotron resonance (TOF-ICR) technique \cite{Gräff:1980,Kšnig199595,brodeur:044318}.
\begin{figure}
\includegraphics[height=.37\textheight]{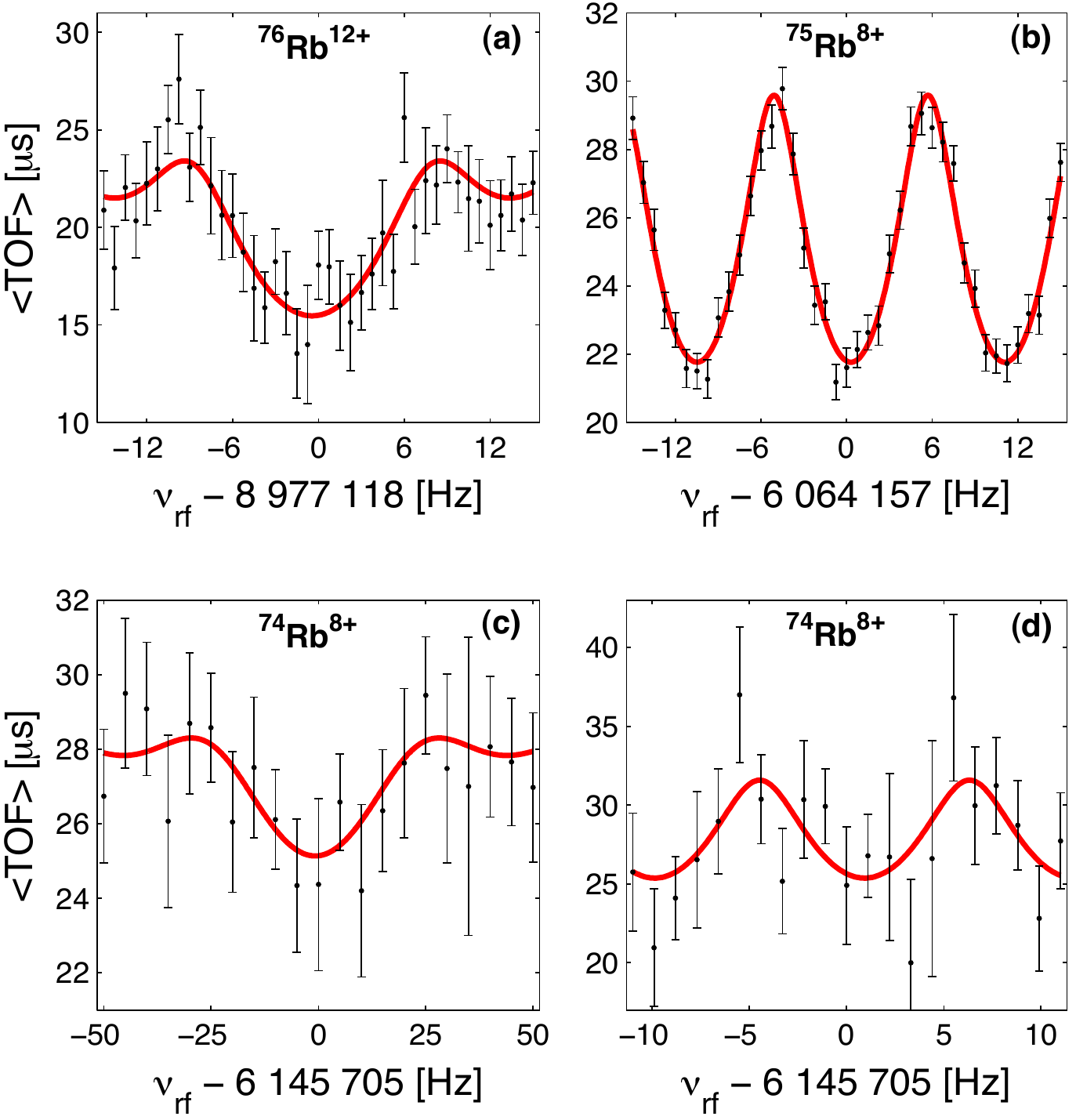}
\caption{\label{fig:resonances}(color online) Time-of-flight ion-cyclotron resonances for Rb-isotopes in charge state $q=+8$ and $q=+12$. During (a) and (c) the rf-field was continuously applied for $T_{\rm rf}=97$ ms and 30 ms, respectively, and a Ramsey excitation scheme with 6-85-6 ms was utilized in (b) and (d). The (red) lines represent the fit to the theoretical line shapes \cite{Kšnig199595,Kretzschmar2007122}.}
\end{figure}
\begin{table}
\caption{\label{tab:MeanRatios} Mean frequency ratios $\overline{R}$ of $^{76,75,74}$Rb$^{q+}$ and $^{74}$Ga$^{8+}$ with respect to $^{85}$Rb$^{9+}$. Where applicable, the ratio and error are the result of a count-class analysis. Uncertainties are displayed as (statistical) and \{stat. + systematics\}.}
\begin{ruledtabular}
\begin{tabular}{lllr}
species&excitation [ms]&$\overline{R}=\nu_c^{\rm ref}/\nu_c^{\rm meas}$  & \#meas\\
\hline
$^{76}$Rb$^{8+}$&97 (conventional)     &1.006067401(15)&5\\
                                 &6-85-6 (Ramsey)&1.006067422(12)&4\\
                                 &previous combined              &1.006067414\{22\}\\
\hline                            
$^{76}$Rb$^{12+}$&97 (conventional)& 0.670692259(23)&1\\
\hline
$^{75}$Rb$^{8+}$&6-85-6 (Ramsey)&0.992864003(10)\{25\}\footnotemark[1]&5\\
\hline
$^{74}$Rb$^{8+}$&30 (conventional)			      &0.979689909(318)\footnotemark[2]&4\\
                                 &6-85-6 (Ramsey)			      &0.979689552(79)\{98\}\footnotemark[2]\footnotemark[3] &3\\
                                 &20(dip)6-65-6(Ram)&0.97968966(10)\{13\}\footnotemark[2]\footnotemark[3]&2\\
                                 &previous combined                                               &0.979689609\{86\}\\	       
\hline
$^{74}$Ga$^{8+}$&97 (conventional)&0.979460129(29)\{71\}\footnotemark[2]&4\\
\end{tabular}
\end{ruledtabular}
\footnotetext[1]{$\nu_c^{\rm meas}$ was determined for $^{75}$Rb$^{8+}$ by a 40-17-40 ms Ramsey scheme, in which the sidebands are less pronounced \cite{Kretzschmar2007122}.}
\footnotetext[2]{Too few ions to perform a count-class analysis.}
\footnotetext[3]{Too few data to study TOF- range dependence. As an upper limit the TOF range dependence for $^{74}$Ga$^{8+}$ was added in quadrature.}
\end{table}Typical time-of-flight ion-cyclotron resonances recorded during these measurements are shown in Fig.\ref{fig:resonances}(a) and (c). At the TOF minimum the rf-frequency $\nu_{\rm rf}$ equals $\nu_{c}$. The width of a resonance, $\Delta \nu_{\rm fwhm}$, is solely governed by the duration $T_{\rm rf}$ of the rf-field \cite{bollen:4355} but is independent of $q$, $m$, or $B$. Hence, the gain in relative precision $\delta m/m= \delta\nu_c/\nu_c$ when utilizing HCI is  due to the larger $\nu_c$ for a given $\Delta \nu_{\rm fwhm}$.  
\newline
As demonstrated recently, a reduction by a factor of 2-3 in $\Delta \nu_{\rm fwhm}$ can be achieved by the Ramsey method of separated oscillatory fields, in which the rf-field  is applied during two pulses separated by a waiting period \cite{PhysRevLett.98.162501,Kretzschmar2007122}. This technique is now used at TITAN, where it can be combined with the advantages offered by HCI. In Fig.\ref{fig:resonances}(b) and (d), a Ramsey excitation scheme with two 6 ms rf-pulses separated by a 85 ms waiting period has been employed (denoted as 6-85-6 ms throughout the paper). With a charge state $q=9+$ in addition to the Ramsey excitation a gain in precision of $\delta\nu_c/\nu_c$ by a factor of about  20 has been achieved compared to the conventional technique. A factor of 36 is possible when optimizing our implementation of the Ramsey technique to the performance reported in \cite{George2007110} and using $q=12+$, the highest charge state for which a measurement has been performed.  
\newline
Due to the $\sqrt{N_{\rm ions}}\cdot q$ dependence of the achievable precision (see Eq. \ref{eq:PenningTrapPrecision}), HCI are favourable as long as the loss in efficiency caused  by the charge breeding is smaller than $q^2$. However, these measurements must be performed with only a few ions (1-5) at a time. If $^{75,76}$Rb had been done with SCI, beam attenuation would have been required. Hence, a lower efficiency for $^{75,76}$Rb$^{+8}$ could be compensated by reducing this beam attenuation. The total number of ions is affected by the breeding in three ways: (1) non-unity efficiency for the chosen charge state due to the charge-state distribution (see Fig.\ref{fig:chargeDistributionTOF}), (2) increased beam emittance causing a reduced transport and trapping efficiency at MPET, and (3) loss of ions due the radioactive decay in the EBIT. For the $^{74}$Rb measurement, ions were kept for 23 ms in the EBIT ($\sim0.35\cdot T_{1/2}$) and the breeding was done in parallel to a measurement in MPET. 
\newline Since HCI are more likely to exchange charge with the residual gas in MPET, the use of HCI demands severe vacuum requirements compared to SCI. When an ion recombines with one or more electrons during the measurement period, its cyclotron frequency changes and the ion is subsequently unaffected by the rf-excitation scheme. These partially recombined ions add to the detected background reducing the sensitivity and possibly induce frequency shifts, e.g. through ion-ion interactions. In preparation for these measurements, we baked the trap, the vacuum vessel, and the extraction beamline. Despite an improved vacuum of $\approx6\cdot10^{-11}$~mbar in the MPET vacuum section, charge exchange occurred, and  in the TOF spectrum we observed an increasing abundance of H$^+_2$ with longer storage time. Even though the current vacuum allowed the recording of a TOF resonance of $^{76}$Rb$^{8+}$ with $T_{\rm rf}=1$ s, ions were typically trapped for 97 ms in MPET as a compromise between increased precision and recombination for longer $T_{\rm rf}$. A summary of the trapping and rf-schemes are listed in Tab.\ref{tab:MeanRatios} together with the measured frequency ratio $R=\nu_c^{\rm ref}/\nu_c^{\rm meas}$ to the reference ion, $^{85}$Rb$^{9+}$. $^{85}$Rb$^{+ }$ was delivered at the same beam energy from TRIUMF's off-line ion source (OLIS) \cite{OLIS}. A fraction of the mass $A=74$ beam from ISAC was $^{74}$Ga,  and we have determined its mass as well. In order to purify the beam and to push the contaminant out of the trap, a dipole rf-field at the reduced cyclotron frequency $\nu_+$ \cite{Blaum20061} of $^{74}$Ga$^{8+}$ was applied for 20 ms during one set of $^{74}$Rb measurements.
\newline
To avoid potential frequency shifts induced by ion-ion interactions, isobaric contamination, or charge exchange, we only considered ion bunches with 5 or less detected ions per detection cycle and performed a count-class analysis \cite{Kellerbauer:EurPhysJD2003} where statistics allowed. For all A=74 measurements, the low count rate made it rare that two or more ions were stored at the same time. To be conservative, we have nevertheless added the difference in $\nu_c$ between 1 and 1-5 detected ions per ion bunch to the systematic uncertainty. Ions of different charge states were produced due to charge exchange in MPET as mentioned above and extracted onto the MPET detection multi-channel plate (MCP), but they could not be resolved in TOF. Thus, the cut on the TOF range was varied in the analysis, and when shifts in $R$ were observed, they were added in quadrature as systematic uncertainties. By far the largest of these shifts was found in $^{74}$Ga, where it accounts for 60 ppb. Other systematic effects due to improper electric field compensation, misalignment between magnetic and trap axes, or harmonic distortions of the electrode structure as well as relativistic effects, were minimized by choosing a reference ion, $^{85}$Rb$^{9+}$, with similar $m/q$. Remaining errors were investigated by measuring $^{85}$Rb$^{10,8+}$ and $^{87}$Rb$^{9+}$ versus $^{85}$Rb$^{9+}$ in various experimental settings. Within one standard deviation, all of these measured $R$ agree with the literature values constraining these systematic uncertainties to  less than 42 ppb for  $^{74}$Rb$^{8+}$ and 20 ppb for the other online measurements with $q=8+$. The different upper limits are  due to a turbo pump failure that necessiated the reconditioning of the electron beam in the EBIT and subsequent retuning of the injection into MPET. According to \cite{PhysRevA.40.6308,Bergstršm2002618}, the image charges do not alter the measurement of $\nu_c$. All systematic uncertainties, including correlations in $R$ due to shared reference frequency measurements, will be discussed in detail in a forthcoming publication.
\begin{figure}
\includegraphics[height=.17\textheight]{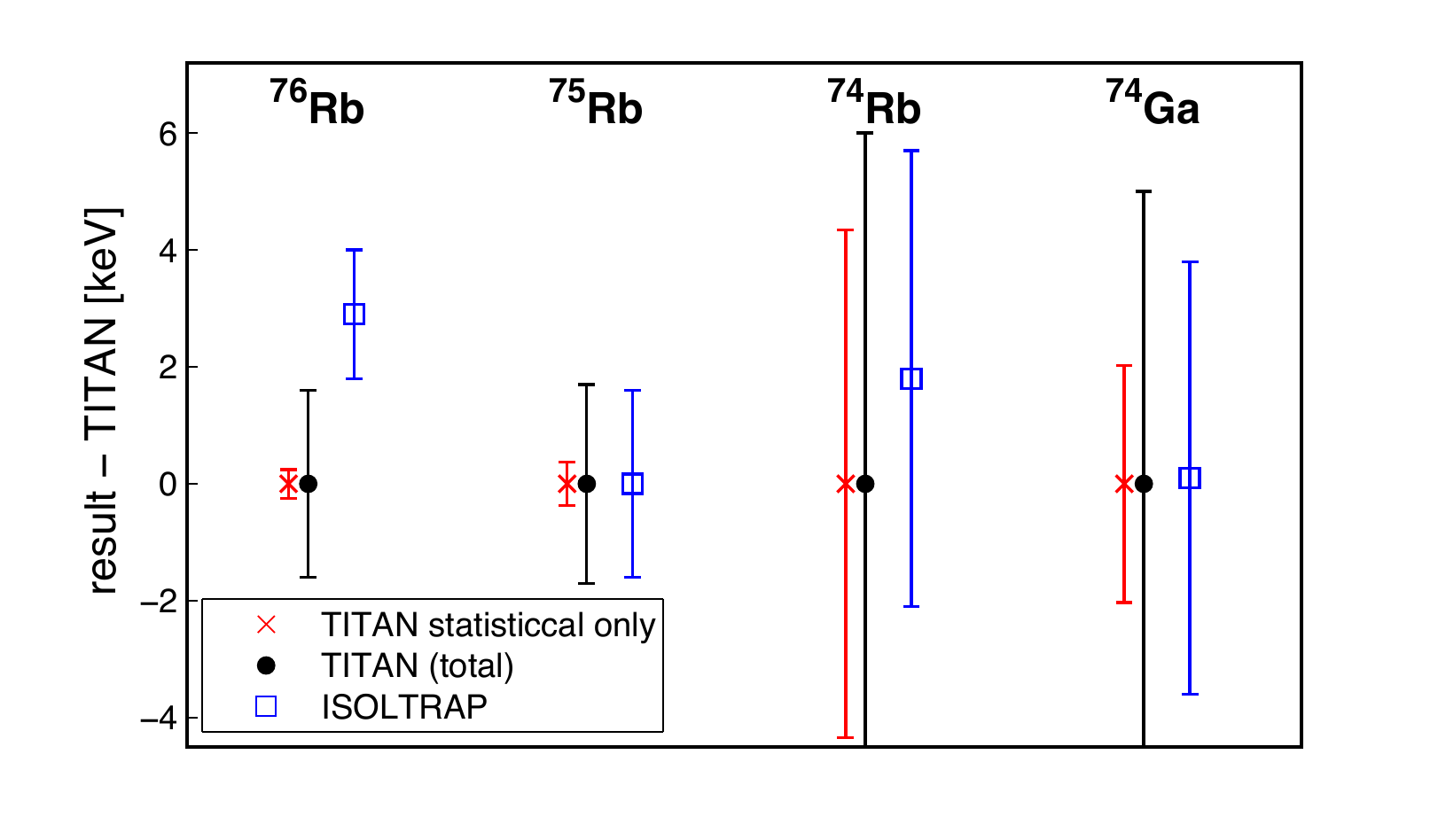}
\caption{\label{fig:results}(color online) Atomic masses of $^{76,75,74}$Rb and $^{74}$Ga in comparison to their respective ISOLTRAP measurements \cite{PhysRevC.76.045504,PhysRevC.75.044303}. Statistical uncertainties are based on fits of ion bunches with 1-5 detected ions but without a count-class analysis}
\end{figure}
\begin{table}
\caption{\label{tab:masses} Mass excess for measured nuclides. $^{76}$Rb$^{12+}$ was a feasibility test  of higher charge
states. Due to its larger systematic uncertainty, only those with $q=8+$ were considered.}
\begin{ruledtabular}
\begin{tabular}{lcc}
nuclide     & this work [keV] & mass evaluation [keV] \\
\hline
$^{76}$Rb&-60481.0(1.6)&-60479.1(0.9)\\
$^{75}$Rb&-57218.7(1.7)&-57218.7(1.2)\\
$^{74}$Rb&-51916.5(6.0)&-51916.0(3.0)\footnotemark[1]\\
$^{74}$Ga&-68049.7(5.0)&-68049.6(3.0)\footnotemark[2]\\
\end{tabular}
\end{ruledtabular}
\footnotetext[1]{\cite{Audi2003337} included a mass value determined from the $\mathcal{F}t$ value of other superallowed $\beta$ emitters and $^{74}$Rb's half-life and branching ratio \cite{PhysRevC.67.051305}. This evaluation does the same with an updated Q-value estimate of  10413.8(7.0) keV \cite{townerPrivateCommunications}  based on \cite{PhysRevC.79.055502}. For the discussion of superallowed decays, we do of course not consider this estimate.}
\footnotetext[2]{Yield measurements determined an isomer to ground state ratio of about 1:190.}
\end{table}
\newline With the measured $\overline{R}$, the atomic masses of the respective nuclides are calculated in Tab. \ref{tab:masses} taking into account the total electron binding energies (0.5, 0.7, and 1.6~keV for Rb$^{8,9,12+}$\cite{10.1063/1.2035727}, respectively, as well as 0.6~keV for Ga$^{8+}$\cite{10.1063/1.2207144}). We also performed a complete atomic mass evaluation based on the procedures in \cite{Audi2003337}, but adding the electron binding energies to the linear equations which were neglected in previous evaluations. Our results are in agreement with ISOLTRAP's measurements \cite{PhysRevC.76.045504,PhysRevC.75.044303} (see Fig.\ref{fig:results}). Due to the use of HCI and the Ramsey excitation they are comparable in precision despite our significantly shorter measurement time ($<$20 h for $^{74}$Rb). Combined with the ISOLTRAP mass value, the total transition energy, $Q_{EC}$, of the superallowed decay in $^{74}$Rb is $10 416.8(3.9)$ keV, an improvement by $\approx12$ \%. This results in a statistical rate function of $f=47283(93)$ \cite{townerPrivateCommunications} and translates together with the recent improvements in $\delta_C$ due the laser spectroscopy work \cite{ManeChargeRadiusRb74} to  a corrected $\mathcal{F}t$-value of 3077(11) s, when considering $\delta_C$ based on shell model calculations with Wood-Saxon radial wavefunctions \cite{PhysRevC.79.055502}. To reduce its uncertainty further, a new mass measurement of $^{74}$Rb and its daughter,  $^{74}$Kr, is planned with charge states up to $q\approx30+$, reachable by a more intense electron beam ($I_e=400$ mA). However, to take full advantage of HCI, it will be necessary to gain better control over systematic effects. Work is underway to improve further the MPET vacuum to avoid charge exchange and its associated uncertainties. It was demonstrated that $m/q$ dependent shifts at TITAN for SCI are at the level of a few ppb \cite{brodeur:044318,MPETsystematics} and an experimental confirmation of this accuracy for HCI is expected. In light of the purly statistical uncertainties of $^{76,75}$Rb in Fig.\ref{fig:results}, such a new measurement could provide knowledge of $^{74}$Rb's Q-value within 0.5 keV. Uncertainties of half-life, branching ratio (BR), or theoretical corrections would then surpass the  $Q_{EC}$-value's contribution to the $\mathcal{F}t$-value's error, stimulating new BR measurements. At this level of experimental precision $f$'s uncertainty in $^{74}$Rb would be dominated by theory and not by the  $Q_{EC}$-value \cite{PhysRevC.71.055501,townerPrivateCommunications}. Perhaps most importantly, more stringent comparisons of conflicting theoretical models of $\delta_c$ similar to \cite{PhysRevC.82.065501} could challenge perceived consistencies between a set of $\delta_C$-calculations, experimental results, and the conserved vector current hypothesis since $^{74}$Rb, with its largest $\delta_C$ among all superallowed $\beta$ emitters, would carry particular weight were it not limited by the current precision in its  $Q_{EC}$-value. 
\newline
In summary, Penning trap mass measurements of highly- charged, short-lived nuclides have been performed for the first time opening a new class of on-line mass measurements with  potentially up to 2 orders of magnitude improved precision versus conventional SCI-TOF-ICR spectroscopy when combined with the Ramsey excitation. This is essential for fundamental symmetries studies, such as presented here for the superallowed $\beta$ emitter  $^{74}$Rb. For nuclear structure and nuclear astrophysics, where the experimental precision is already sufficient, this novel technique will reduce the measurement time and thus allow one to map the nuclear mass landscape more quickly. In addition, the same precision can be achieved for lower production yields and/or shorter half-lives. At this time, TITAN is unique in providing the possibility for high charge states in Penning trap mass spectroscopy of radionuclides.
\newline
This work has been supported by the Natural Sciences and Engineering Research Council of Canada and the National Research Council of Canada. We would like to thank the TRIUMF technical staff, especially M. Good. We are very grateful to I.S. Towner for discussions and his calculations. S.E. acknowledges support from the Vanier CGS , T.B. from the Evangelisches Studienwerk e.V. Villigst, and V. V. S. from the Deutsche Studienstiftung.


\begin{thebibliography}{39}
\expandafter\ifx\csname natexlab\endcsname\relax\def\natexlab#1{#1}\fi
\expandafter\ifx\csname bibnamefont\endcsname\relax
  \def\bibnamefont#1{#1}\fi
\expandafter\ifx\csname bibfnamefont\endcsname\relax
  \def\bibfnamefont#1{#1}\fi
\expandafter\ifx\csname citenamefont\endcsname\relax
  \def\citenamefont#1{#1}\fi
\expandafter\ifx\csname url\endcsname\relax
  \def\url#1{\texttt{#1}}\fi
\expandafter\ifx\csname urlprefix\endcsname\relax\def\urlprefix{URL }\fi
\providecommand{\bibinfo}[2]{#2}
\providecommand{\eprint}[2][]{\url{#2}}

\bibitem[{\citenamefont{Bollen et~al.}(1987)}]{springerlink:10.1007/BF02394875}
\bibinfo{author}{\bibfnamefont{G.}~\bibnamefont{Bollen}} \bibnamefont{et~al.},
  \bibinfo{journal}{Hyperfine Interactions} \textbf{\bibinfo{volume}{38}},
  \bibinfo{pages}{793} (\bibinfo{year}{1987}).

\bibitem[{\citenamefont{Stolzenberg et~al.}(1990)}]{PhysRevLett.65.3104}
\bibinfo{author}{\bibfnamefont{H.}~\bibnamefont{Stolzenberg}}
  \bibnamefont{et~al.}, \bibinfo{journal}{Phys. Rev. Lett.}
  \textbf{\bibinfo{volume}{65}}, \bibinfo{pages}{3104} (\bibinfo{year}{1990}).

\bibitem[{\citenamefont{Blaum}(2006)}]{Blaum20061}
\bibinfo{author}{\bibfnamefont{K.}~\bibnamefont{Blaum}},
  \bibinfo{journal}{Physics Reports} \textbf{\bibinfo{volume}{425}},
  \bibinfo{pages}{1 } (\bibinfo{year}{2006}).

\bibitem[{\citenamefont{Blaum and
  Block}(2009)}]{springerlink:10.1007/s10751-009-0031-8}
\bibinfo{author}{\bibfnamefont{K.}~\bibnamefont{Blaum}} \bibnamefont{and}
  \bibinfo{author}{\bibfnamefont{M.}~\bibnamefont{Block}},
  \bibinfo{journal}{Hyp. Int.} \textbf{\bibinfo{volume}{194}},
  \bibinfo{pages}{65} (\bibinfo{year}{2009}).

\bibitem[{\citenamefont{Smith et~al.}(2008)}]{PhysRevLett.101.202501}
\bibinfo{author}{\bibfnamefont{M.}~\bibnamefont{Smith}} \bibnamefont{et~al.},
  \bibinfo{journal}{Phys. Rev. Lett.} \textbf{\bibinfo{volume}{101}},
  \bibinfo{pages}{202501} (\bibinfo{year}{2008}).

\bibitem[{\citenamefont{Block et~al.}(2010)}]{BlockNature463.785}
\bibinfo{author}{\bibfnamefont{M.}~\bibnamefont{Block}} \bibnamefont{et~al.},
  \bibinfo{journal}{Nature} \textbf{\bibinfo{volume}{463}},
  \bibinfo{pages}{785} (\bibinfo{year}{2010}).

\bibitem[{\citenamefont{Bollen}(2001)}]{Bollen20013}
\bibinfo{author}{\bibfnamefont{G.}~\bibnamefont{Bollen}},
  \bibinfo{journal}{Nuclear Physics A} \textbf{\bibinfo{volume}{693}},
  \bibinfo{pages}{3 } (\bibinfo{year}{2001}).

\bibitem[{\citenamefont{Carlberg et~al.}(1999)}]{PhysRevLett.83.4506}
\bibinfo{author}{\bibfnamefont{C.}~\bibnamefont{Carlberg}}
  \bibnamefont{et~al.}, \bibinfo{journal}{Phys. Rev. Lett.}
  \textbf{\bibinfo{volume}{83}}, \bibinfo{pages}{4506} (\bibinfo{year}{1999}).

\bibitem[{\citenamefont{Herfurth
  et~al.}(2002)}]{springerlink:10.1140/epja/i2001-10216-x}
\bibinfo{author}{\bibfnamefont{F.}~\bibnamefont{Herfurth}}
  \bibnamefont{et~al.}, \bibinfo{journal}{EPJ A} \textbf{\bibinfo{volume}{15}},
  \bibinfo{pages}{17} (\bibinfo{year}{2002}).

\bibitem[{\citenamefont{Kellerbauer et~al.}(2004)}]{PhysRevLett.93.072502}
\bibinfo{author}{\bibfnamefont{A.}~\bibnamefont{Kellerbauer}}
  \bibnamefont{et~al.}, \bibinfo{journal}{Phys. Rev. Lett.}
  \textbf{\bibinfo{volume}{93}}, \bibinfo{pages}{072502}
  (\bibinfo{year}{2004}).

\bibitem[{\citenamefont{Kellerbauer et~al.}(2007)}]{PhysRevC.76.045504}
\bibinfo{author}{\bibfnamefont{A.}~\bibnamefont{Kellerbauer}}
  \bibnamefont{et~al.}, \bibinfo{journal}{Phys. Rev. C}
  \textbf{\bibinfo{volume}{76}}, \bibinfo{pages}{045504}
  (\bibinfo{year}{2007}).

\bibitem[{\citenamefont{Hardy and Towner}(2009)}]{PhysRevC.79.055502}
\bibinfo{author}{\bibfnamefont{J.~C.} \bibnamefont{Hardy}} \bibnamefont{and}
  \bibinfo{author}{\bibfnamefont{I.~S.} \bibnamefont{Towner}},
  \bibinfo{journal}{PRC} \textbf{\bibinfo{volume}{79}}, \bibinfo{pages}{055502}
  (\bibinfo{year}{2009}).

\bibitem[{\citenamefont{Man\'{e} et~al.}(2011)}]{ManeChargeRadiusRb74}
\bibinfo{author}{\bibfnamefont{E.}~\bibnamefont{Man\'{e}}}
  \bibnamefont{et~al.}, \bibinfo{journal}{submitted to Phys. Rev. Lett.}
  (\bibinfo{year}{2011}).

\bibitem[{\citenamefont{Grinyer et~al.}(2010)}]{Grinyer2010236}
\bibinfo{author}{\bibfnamefont{G.~F.} \bibnamefont{Grinyer}}
  \bibnamefont{et~al.}, \bibinfo{journal}{Nucl. Instr. Meth. A}
  \textbf{\bibinfo{volume}{622}}, \bibinfo{pages}{236 } (\bibinfo{year}{2010}).

\bibitem[{\citenamefont{Towner and Hardy}(2010)}]{PhysRevC.82.065501}
\bibinfo{author}{\bibfnamefont{I.~S.} \bibnamefont{Towner}} \bibnamefont{and}
  \bibinfo{author}{\bibfnamefont{J.~C.} \bibnamefont{Hardy}},
  \bibinfo{journal}{PRC} \textbf{\bibinfo{volume}{82}}, \bibinfo{pages}{065501}
  (\bibinfo{year}{2010}).

\bibitem[{\citenamefont{Dilling et~al.}(2006)}]{Dilling2006198}
\bibinfo{author}{\bibfnamefont{J.}~\bibnamefont{Dilling}} \bibnamefont{et~al.},
  \bibinfo{journal}{IJMS} \textbf{\bibinfo{volume}{251}}, \bibinfo{pages}{198 }
  (\bibinfo{year}{2006}).

\bibitem[{\citenamefont{Dombsky et~al.}(2000)}]{dombsky:978}
\bibinfo{author}{\bibfnamefont{M.}~\bibnamefont{Dombsky}} \bibnamefont{et~al.},
  \bibinfo{journal}{Rev. Sci. Instrum.} \textbf{\bibinfo{volume}{71}},
  \bibinfo{pages}{978 } (\bibinfo{year}{2000}).

\bibitem[{\citenamefont{Brunner et~al.}(2011{\natexlab{a}})}]{TITAN-RFQ}
\bibinfo{author}{\bibfnamefont{T.}~\bibnamefont{Brunner}} \bibnamefont{et~al.},
  \bibinfo{journal}{arXiv:1107.2187}  (\bibinfo{year}{2011}{\natexlab{a}}).

\bibitem[{\citenamefont{Lapierre et~al.}(2010)}]{Lapierre201054}
\bibinfo{author}{\bibfnamefont{A.}~\bibnamefont{Lapierre}}
  \bibnamefont{et~al.}, \bibinfo{journal}{Nucl. Instr. Meth. A}
  \textbf{\bibinfo{volume}{624}}, \bibinfo{pages}{54 } (\bibinfo{year}{2010}).

\bibitem[{\citenamefont{Brunner et~al.}(2011{\natexlab{b}})}]{EXO-TITAN-BNgate}
\bibinfo{author}{\bibfnamefont{T.}~\bibnamefont{Brunner}} \bibnamefont{et~al.},
  \bibinfo{journal}{arXiv:1107.4010, accepted for publication in IJMS}
  (\bibinfo{year}{2011}{\natexlab{b}}).

\bibitem[{\citenamefont{Gr{\"{a}}ff et~al.}(1980)}]{Gräff:1980}
\bibinfo{author}{\bibfnamefont{G.}~\bibnamefont{Gr{\"{a}}ff}}
  \bibnamefont{et~al.}, \bibinfo{journal}{Zeitschrift f{\"{u}}r Physik A}
  \textbf{\bibinfo{volume}{297}}, \bibinfo{pages}{35} (\bibinfo{year}{1980}).

\bibitem[{\citenamefont{K{\"{o}}nig et~al.}(1995)}]{Kšnig199595}
\bibinfo{author}{\bibfnamefont{M.}~\bibnamefont{K{\"{o}}nig}}
  \bibnamefont{et~al.}, \bibinfo{journal}{IJMS} \textbf{\bibinfo{volume}{142}},
  \bibinfo{pages}{95 } (\bibinfo{year}{1995}).

\bibitem[{\citenamefont{Brodeur et~al.}(2009)}]{brodeur:044318}
\bibinfo{author}{\bibfnamefont{M.}~\bibnamefont{Brodeur}} \bibnamefont{et~al.},
  \bibinfo{journal}{Physical Review C} \textbf{\bibinfo{volume}{80}},
  \bibinfo{pages}{044318} (\bibinfo{year}{2009}).

\bibitem[{\citenamefont{Kretzschmar}(2007)}]{Kretzschmar2007122}
\bibinfo{author}{\bibfnamefont{M.}~\bibnamefont{Kretzschmar}},
  \bibinfo{journal}{IJMS} \textbf{\bibinfo{volume}{264}}, \bibinfo{pages}{122 }
  (\bibinfo{year}{2007}).

\bibitem[{\citenamefont{Bollen et~al.}(1990)}]{bollen:4355}
\bibinfo{author}{\bibfnamefont{G.}~\bibnamefont{Bollen}} \bibnamefont{et~al.},
  \bibinfo{journal}{J. App.Phys.} \textbf{\bibinfo{volume}{68}},
  \bibinfo{pages}{4355} (\bibinfo{year}{1990}).

\bibitem[{\citenamefont{George
  et~al.}(2007{\natexlab{a}})}]{PhysRevLett.98.162501}
\bibinfo{author}{\bibfnamefont{S.}~\bibnamefont{George}} \bibnamefont{et~al.},
  \bibinfo{journal}{Phys. Rev. Lett.} \textbf{\bibinfo{volume}{98}},
  \bibinfo{pages}{162501} (\bibinfo{year}{2007}{\natexlab{a}}).

\bibitem[{\citenamefont{George et~al.}(2007{\natexlab{b}})}]{George2007110}
\bibinfo{author}{\bibfnamefont{S.}~\bibnamefont{George}} \bibnamefont{et~al.},
  \bibinfo{journal}{IJMS} \textbf{\bibinfo{volume}{264}}, \bibinfo{pages}{110 }
  (\bibinfo{year}{2007}{\natexlab{b}}).

\bibitem[{\citenamefont{Jayamanna et~al.}(2008)}]{OLIS}
\bibinfo{author}{\bibfnamefont{K.}~\bibnamefont{Jayamanna}}
  \bibnamefont{et~al.}, \bibinfo{journal}{Rev. Sci. Instr.}
  \textbf{\bibinfo{volume}{79}}, \bibinfo{pages}{02C711}
  (\bibinfo{year}{2008}).

\bibitem[{\citenamefont{{A. Kellerbauer}
  et~al.}(2003)}]{Kellerbauer:EurPhysJD2003}
\bibinfo{author}{\bibnamefont{{A. Kellerbauer}}} \bibnamefont{et~al.},
  \bibinfo{journal}{Eur. Phys. J. D} \textbf{\bibinfo{volume}{22}},
  \bibinfo{pages}{53} (\bibinfo{year}{2003}).

\bibitem[{\citenamefont{Van~Dyck et~al.}(1989)}]{PhysRevA.40.6308}
\bibinfo{author}{\bibfnamefont{R.~S.} \bibnamefont{Van~Dyck}}
  \bibnamefont{et~al.}, \bibinfo{journal}{Phys. Rev. A}
  \textbf{\bibinfo{volume}{40}}, \bibinfo{pages}{6308} (\bibinfo{year}{1989}).

\bibitem[{\citenamefont{Bergstr{\"o}m et~al.}(2002)}]{Bergstršm2002618}
\bibinfo{author}{\bibfnamefont{I.}~\bibnamefont{Bergstr{\"o}m}}
  \bibnamefont{et~al.}, \bibinfo{journal}{NIM A}
  \textbf{\bibinfo{volume}{487}}, \bibinfo{pages}{618 } (\bibinfo{year}{2002}).

\bibitem[{\citenamefont{Gu\'enaut et~al.}(2007)}]{PhysRevC.75.044303}
\bibinfo{author}{\bibfnamefont{C.}~\bibnamefont{Gu\'enaut}}
  \bibnamefont{et~al.}, \bibinfo{journal}{Phys. Rev. C}
  \textbf{\bibinfo{volume}{75}}, \bibinfo{pages}{044303}
  (\bibinfo{year}{2007}).

\bibitem[{\citenamefont{Audi et~al.}(2003)}]{Audi2003337}
\bibinfo{author}{\bibfnamefont{G.}~\bibnamefont{Audi}} \bibnamefont{et~al.},
  \bibinfo{journal}{Nuclear Physics A} \textbf{\bibinfo{volume}{729}},
  \bibinfo{pages}{337 } (\bibinfo{year}{2003}).

\bibitem[{\citenamefont{Piechaczek et~al.}(2003)}]{PhysRevC.67.051305}
\bibinfo{author}{\bibfnamefont{A.}~\bibnamefont{Piechaczek}}
  \bibnamefont{et~al.}, \bibinfo{journal}{Phys. Rev. C}
  \textbf{\bibinfo{volume}{67}}, \bibinfo{pages}{051305(R)}
  (\bibinfo{year}{2003}).

\bibitem[{\citenamefont{Towner}(2011)}]{townerPrivateCommunications}
\bibinfo{author}{\bibfnamefont{I.~S.} \bibnamefont{Towner}},
  \bibinfo{journal}{private communications}  (\bibinfo{year}{2011}).

\bibitem[{\citenamefont{Sansonetti}(2006)}]{10.1063/1.2035727}
\bibinfo{author}{\bibfnamefont{J.~E.} \bibnamefont{Sansonetti}},
  \bibinfo{journal}{J.Phys.Chem.Ref.Data} \textbf{\bibinfo{volume}{35}},
  \bibinfo{pages}{301} (\bibinfo{year}{2006}).

\bibitem[{\citenamefont{Shirai et~al.}(2007)}]{10.1063/1.2207144}
\bibinfo{author}{\bibfnamefont{T.}~\bibnamefont{Shirai}} \bibnamefont{et~al.},
  \bibinfo{journal}{J.Phys.Chem.Ref. Data} \textbf{\bibinfo{volume}{36}},
  \bibinfo{pages}{509} (\bibinfo{year}{2007}).

\bibitem[{\citenamefont{Brodeur et~al.}(2011)}]{MPETsystematics}
\bibinfo{author}{\bibfnamefont{M.}~\bibnamefont{Brodeur}} \bibnamefont{et~al.},
  \bibinfo{journal}{submitted to to IJMS}  (\bibinfo{year}{2011}).

\bibitem[{\citenamefont{Hardy and Towner}(2005)}]{PhysRevC.71.055501}
\bibinfo{author}{\bibfnamefont{J.~C.} \bibnamefont{Hardy}} \bibnamefont{and}
  \bibinfo{author}{\bibfnamefont{I.~S.} \bibnamefont{Towner}},
  \bibinfo{journal}{PRC} \textbf{\bibinfo{volume}{71}}, \bibinfo{pages}{055501}
  (\bibinfo{year}{2005}).

\end{thebibliography}
\end{document}